\documentclass[preprint,12pt,english]{elsarticle}
\usepackage[english]{babel}
\usepackage{multicol}
\usepackage{amsmath,amsthm,amsfonts,amssymb}
\usepackage{dsfont}
\usepackage{adjustbox}
\usepackage{graphicx}
\usepackage{latexsym}
\usepackage{epsfig}
\biboptions{square,numbers,sort&compress}
\usepackage{xcolor}
\usepackage{hyperref}
\hypersetup{
	colorlinks,
    citecolor=blue!20!green!90!black!99,
    filecolor=red,
    linkcolor=blue!20!red!90!black!99,
    urlcolor=gray
    }

\begin{document}
\begin{frontmatter}

\journal{Physics Letters B}

\title{New class of solutions of a generalized $O(3)$-sigma Chern-Simons model}

\author[pici]{F. C. E. Lima}
\ead{cleiton.estevao@fisica.ufc.br}

\author[pici]{D. M. Dantas}
\ead{davi@fisica.ufc.br}

\author[pici]{C. A. S. Almeida}
\ead{carlos@fisica.ufc.br}

\address[pici]{Universidade Federal do Cear\'a (UFC), Departamento de F\'isica, Campus do Pici, Fortaleza - CE, C.P. 6030, 60455-760 - Brazil}

\begin{keyword}
$O(3)$-sigma Model \sep Chern-Simons Term \sep Compacton-like configuration.
\end{keyword}

\begin{abstract}
In this work, we investigated the existence of compacton-like configuration in the $O(3)$-sigma model. We consider a minimally coupled $O(3)$-sigma model with a gauge field governed by a generalized Chern-Simons term. Contrary to that established in the literature, we impose a new set of boundary conditions and, we find solutions of the variable fields and the respective energy density in the Bogomol'nyi limit. On the other hand, the introduction of a parameter $\omega$ in the Chern-Simons term can be adjusted to leads to finite-energy solutions of the model. Moreover, compact-like structures were studied with the evolution of this $\omega$ generalized Chern-Simons term.
\end{abstract}
\end{frontmatter}

\section{Introduction}

In 1993, compactons were defined as finite wavelength solitons \cite{ref-compactons}, and so far they have been the object of several studies, since models containing topological defects were used to represent particles and cosmological objects, such as cosmic strings \cite{2}. A particular arrangement of particles can be represented as a group of compactons, and in this case, we will not have the problem of particle overlap, or defects \cite{3}. Another reason for the growing interest in the study of compact-like structures is compact vortices and the so-called skyrmions, that are intrinsically connected with recent advances in nanometer-scale material miniaturization for spintronic applications \cite{4, 5}. Moreover, the quantum compacton properties were studied in the reference \cite{quantum-compactons}. Features of compacton models at finite temperature were discussed in Ref. \cite{temperature} and the applications of compactons on braneworlds context in Refs. \cite{6, 6b, 6c}.

The solitons solutions of the gauged $O(3)$-sigma model are relevant in several systems, among them,  condensed matter systems \cite{7, 8}. In a version of this model with the Maxwell term, self-dual solutions were discussed in Ref. \cite{9}. In other similar models such as the Skyrme model, a symmetry gauge is required where we can use a scale invariance breaking of a $O(3)$-sigma model with the field dynamics governed by a term a Maxwell term, where the gauge group $U(1)$ is a subgroup of $O(3)$ \cite{Casana-Skyrme}. On the other hand, models in which a Chern-Simons term governs the dynamics of the gauge field, have been studied in other works \cite{10, 11, 12, bazeia}.

In this work, we will study the existence of soliton-like solutions in the (2+1)D $O(3)$-sigma model with the gauge field governed by a generalized Chern-Simons term. Therefore, in section \ref{section2} we started discussing the dynamics gauge field for this generalized model. In section \ref{section3}, we began our investigation of the suitable solutions of the model. In order to accomplish this, we use a new set of boundary conditions to the variable field, namely, $f(r=0)=0$ and $f(r\rightarrow\infty)=\pi$. Starting from this premise, in the section \ref{section4}, we studied the possibility of solutions with finite energy and the existence of compact-type defects. Our discussions and conclusions are presented in Section \ref{section5}.

\section{The generalized $O(3)$-sigma Chern-Simons model}\label{section2}

To construct the model, we consider combinations of Lagrangian densities described by references \cite{3} and \cite{12} (we work in natural unity $c=\hbar=1$). Namely,
\begin{equation}\label{lagrang}
\mathcal{L}=\frac{1}{2}D_{\mu}\vec{\phi}\cdot D^{\mu}\vec{\phi}+\frac{\kappa}{4}\omega\varepsilon^{\mu\nu\rho}A_{\mu}F_{\nu\rho}-\frac{1}{2\kappa^{2}}(1+\hat{n}_{3}\cdot \vec{\phi})(1-\hat{n}_{3}\cdot \vec{\phi})^{3}.
\end{equation}
%The function $G(\phi)$ multiplying the kinetic term for scalar fields was introduced by M. A. Lohe \cite{lohe}, reassessed years later by Y. Verdin \textit{et al.} \cite{verdin} and recently used by some other authors \cite{3, bazeia}.

As we will show below, the adimensional parameter $\omega$ multiplying the CS term, is responsible for bringing finitude to the energy density in this generalized model. Besides {$\vec{\phi}=\phi_1\hat{n}_{1}+\phi_2\hat{n}_{2}+\phi_3\hat{n}_{3}$} is a vector with three components and norm $\vec{\phi}\cdot\vec{\phi}=1$ \cite{12,13,14}. In this context, we consider the Minkowski flat-space signature $g_{\mu\nu}=(+,-,-)$. Moreover, the particular choice of $\omega$  \cite{bazeia,3} allows the existence of compacton-like structures to the $O(3)$ sigma model. The electromagnetic tensor reads $F_{\nu\rho}=\partial_{\nu} A_{\rho}-\partial_{\rho} A_{\nu}$, where $A_{\mu}$ represents the gauge field and the $\varepsilon^{\mu\nu\rho}$ is the Levi-Civita symbol.

The covariant derivative is defined as \cite{12} 
\begin{equation}
D_{\mu}\vec{\phi}=\partial_{\mu}\vec{\phi}+A_{\mu}\hat{n}_{3}\times\vec{\phi}
\end{equation}

The Lagrangian (\ref{lagrang}) is invariant under a $SO(2)$ isorotation around the axis $\hat{n}_{3}$. In fact, one can  use the identity $D_{\mu}\vec{\phi}\cdot D^{\mu}\vec{\phi}= |(\partial_{\mu}+iA_{\mu})(\phi_{1}+i\phi_{2})|^{2}+$\newline $\partial_{\mu}\phi_{3}\partial^{\mu}\phi_{3}$ to see the local  $U(1)$ nature of it. From the expression (\ref{lagrang}) we see that the potential has two degenerate minima at $\phi_{3}=\pm 1$. The constraint $\vec{\phi}\cdot\vec{\phi}=1$ implies the $\phi_{1}$ and $\phi_{2}$ are zero when $\phi_{3}=\pm 1$.

The equations of the motion for the model is \cite{12}
\begin{equation}
D_{\mu}{\bf J}^{\mu}=-\frac{1}{\kappa^{2}}(\hat{n}_{3}\times\vec{\phi})(1-\hat{n}_{3}\cdot\vec{\phi})^{2}(1+2\hat{n}_{3}\cdot \vec{\phi}),
\end{equation}
where
\begin{equation}
j^{\mu}=\frac{\kappa}{2}\omega\varepsilon^{\mu\nu\rho}F_{\nu\rho},
\end{equation}
and the current ${\bf J}^{\mu}$ is defined as 
\begin{equation}
{\bf J}^{\mu}= \vec{\phi}\times D^{\mu}\vec{\phi},
\end{equation}
where the $U(1)$ current is $j^{\mu}=-{\bf J}^{\mu}\cdot \hat{n}_{3}$.

Building the functional of energy, we obtain the expression
\begin{align}\label{energy}
E=& \frac{1}{2}\int d^2x \, \bigg[(D_{1}\vec{\phi})^{2}+(D_{2}\vec{\phi})^{2}+\frac{\omega^{2}\kappa^{2}F^{2}_{12}}{\phi_{1}^{2}+\phi_{2}^{2}}+ \frac{1}{\kappa^{2}}(1+\hat{n}_{3}\cdot \vec{\phi})(1-\hat{n}_{3}\cdot \vec{\phi})^{3}\bigg]
\end{align}

The potential $A_0$ has been eliminated by using the Gauss law \cite{15}. The energy functional can be {rearranged} as 
\begin{equation}\label{energy1}
\begin{split}
E=&\int d^2x \; \bigg[(D_{i}\vec{\phi}\pm \varepsilon_{ij}\vec{\phi}\times D_{j}\vec{\phi})^{2}+ \frac{\omega^{2}\kappa^{2}}{1-\phi_{3}^{2}}\times \bigg(F_{12}\mp\frac{1}{\omega\kappa^{2}}(1+\phi_{3})(1-\phi_{3})^{2}\bigg)^{2} \; \bigg]\\ &\pm 4\pi\int d^{2}x  K_{0},
\end{split}
\end{equation}
where $K_{0}$ is the zero component of the topological current, and $K_{\mu}$ is defined as 
\begin{equation}
K_{\mu}=\frac{1}{8\pi}\varepsilon_{\mu\nu\rho}\bigg[\vec{\phi}\cdot D^{\mu}\vec{\phi}\times D^{\rho}\vec{\phi}+\omega F^{\nu\rho}(1-\hat{n}_{3}\cdot \vec{\phi})\bigg].
\end{equation}

Applying the BPS limit in the expression (\ref{energy1}) we get 
\begin{equation}
 D_{i}\vec{\phi}\pm \varepsilon_{ij}\, \vec{\phi}\times D_{j}\vec{\phi}=0,
\end{equation}
\begin{equation}
F_{12}\mp \frac{1}{\omega\kappa^{2}}(1+\phi_{3})(1-\phi_{3})^{2}=0.
\end{equation}

At the BPS limit the energy density of the model is reduced to
\begin{equation}\label{energy-bps}
\mathcal{E}_{BPS}=\frac{1}{2}\vert D_{i}\phi\vert^{2}+F_{12}(1-\hat{n}\cdot\phi).
\end{equation}

\section{The solutions of the model}\label{section3}
In order to study the topological solutions of the model, we chose an ansatz for the field variables. Our choice is the same as Gosh \cite{12}, namely
\begin{equation}\label{ansatz}
\phi_{1}(\rho, \theta)=\sin f(r)\cos N\theta;
\end{equation}
\begin{equation}\label{ansatz1}
\phi_{2}(\rho, \theta)=\sin f(r)\cos N\theta;
\end{equation}
\begin{equation}\label{ansatz2}
{\bf \phi}_{3}(\rho, \theta)=\cos f(r).
\end{equation}

We have also chosen
\begin{equation}\label{ansatz3}
\vec{A}=-\hat{e}_{\theta}\frac{Na(r)}{\kappa r}.
\end{equation}

We highlight that $f(r)$ is an arbitrary dimensionless function ($r=\rho/\kappa$). $N$ is an integer and also defines the degree of a topological soliton.

Replacing the expressions (\ref{ansatz}, \ref{ansatz1}, \ref{ansatz2} and \ref{ansatz3}) in the Bogolmo'nyi equations of the model, we get 
\begin{equation}\label{1}
f'(r)=\pm 2N\frac{a+1}{r}\sin\frac{f(r)}{2}\cos\frac{f(r)}{2}
\end{equation}
and
\begin{equation}\label{2}
a'(r)=\pm\frac{2r}{N\omega}\sin^{2} f(r)\sin^{2}\frac{f(r)}{2}.
\end{equation}

Decoupling the above equations, we obtain
\begin{equation}\label{equationself}
f''(r)-\frac{f'(r)^{2}}{\tan f(r)}+\frac{f'(r)}{r}-\frac{2}{\omega}\sin^{3}f(r)\sin^{2}\frac{f(r)}{2}=0.
\end{equation}

Now, we proceed to investigate the effects of the coupled parameter to the  Chern-Simons field to the usual gauged $O(3)$ sigma model.
%It is worth noting that the coupling term $G$ does not change the shape of the solutions of the variable field, so we need only investigate the effects of coupling term $\omega$  to the usual gauged $O(3)$ sigma model.

The equations (\ref{1}) and (\ref{2}) are invariant under the transformation $f(r)\rightarrow f(r)+2\pi$. This observation is sufficient to motivate the study of the above equations with $f(r)$ having any value between $0$ to $2\pi$.

As suggested by Gosh \cite{12}, we can conveniently take the change in the dependent variable as
\begin{equation}\label{trans1}
\eta_{1}=\pi+f(r)
\end{equation}
or 
\begin{equation}\label{trans2}
\eta_{2}=\pi-f(r)
\end{equation}
where we consider that $a(r)$ remains unchanged.

We note that the previous transformations maintain eqs. (\ref{1}) and (\ref{2}) invariants. We conclude that for a particular profile of $a(r)$ the solution for $f(r)$ is symmetric about $f(r)=\pi$. Thus we can restrict the asymptotic values of $f(r)$ between $0$ and $\pi$. Therefore, once we get the solution in this interval, there must be a symmetric solution in the interval of $\pi$ to $2\pi$.

%Although some authors \cite{12} assert that to getting of finite energy solutions the variable field (near the origin) depends of the conditions $f(0)=\pi$ and $a(0)=0$, in the case not generalized, we will consider the study numerical of the model generalized for the conditions $f(0)=0$ and $f(\infty)=\pi$. It is necessary to mention that when $f(r\rightarrow\infty) = \pi$, the topological charge of the model is not an integer. Therefore, no topological argument can be used for the stability of the solutions with these boundary conditions.
Although some authors assert that to getting of finite energy solutions in the case not generalized depends on the conditions $f(0)=\pi$ and $a(0)=0$, we will consider the numerical analysis with the conditions $f(0)=0$ and $f(\infty)=\pi$.  It is worthwhile to mention that when $f(r\rightarrow\infty)=\pi$, the topological charge of the model is an integer. Therefore, topological arguments similar to that used by Gosh can be used for discuss the stability of the solutions with these boundary conditions. For more detailed and comprehensive reading, we recommend the papers \cite{9,12}.

\section{Numerical results}\label{section4}

In this section, we start the numerical study of the self-dual equations for the sigma model with generalized Chern-Simons term. Let us now turn our attention to the statement proposed in Ref. \cite{12}. In this work, Ghosh claims that solutions with physical interest are solutions of localized energy. To obtain these solutions, we must analyse the existence of solitons in a specified contour. For the nontopological case the contour is $f(r=0)=\pi$ and $f(r\rightarrow\infty)=\pi$. For the topological solitons we must consider $f(r=0)=\pi$ and $f(r\rightarrow\infty)=0$.

\begin{figure}[!htb]
\centering
\includegraphics[scale=0.85]{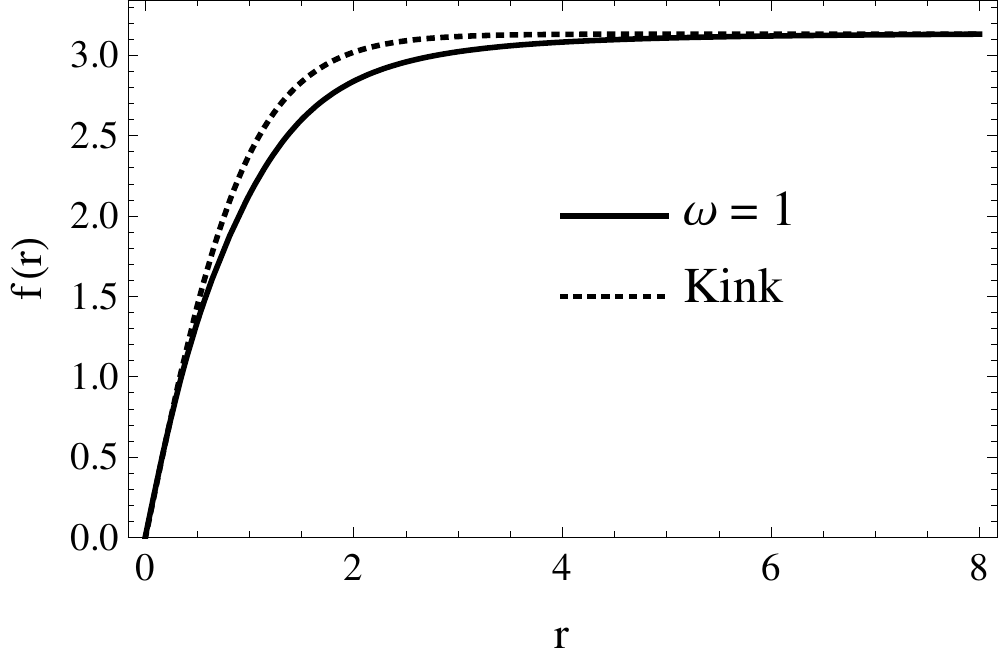}
\caption{Non-topological solitons for $f(r=0)=0$ and $f(r\rightarrow\infty) =\pi $.}
\label{fig-1}
\end{figure}

However, adopting the condition $f(r=0)=0$ (which was not studied in Ref. \cite{12}), we obtain a kink-like solution with localized energy. In order to demonstrate it, we return to equation (\ref{equationself}) and analyse the existence of such solutions, through a interpolation method. We consider that the variable field near the origin behaves as $f(r)=a_{0}r^{N}$. The solution for $f(r)$ is presented in Fig. \ref{fig-1}.

%In contradistinction to the work proposed by Ghosh, we will start the investigation for the behavior of the variable field being initially $f(r=0)=0$, we show that despite the work presented in the literature, the existence of solutions with localized energy, presents Kink-like solutions for this contour. To show this, we return the equation (\ref{equationself}) and analyze, through the interpolation method, the existence of such solutions, for this, we consider that the variable field near of the origin behaves as $f(r)=a_{0}r^{N}$. As a result we have the graphic solutions presented below.

\begin{figure}[!htb]
\centering
\includegraphics[scale=0.85]{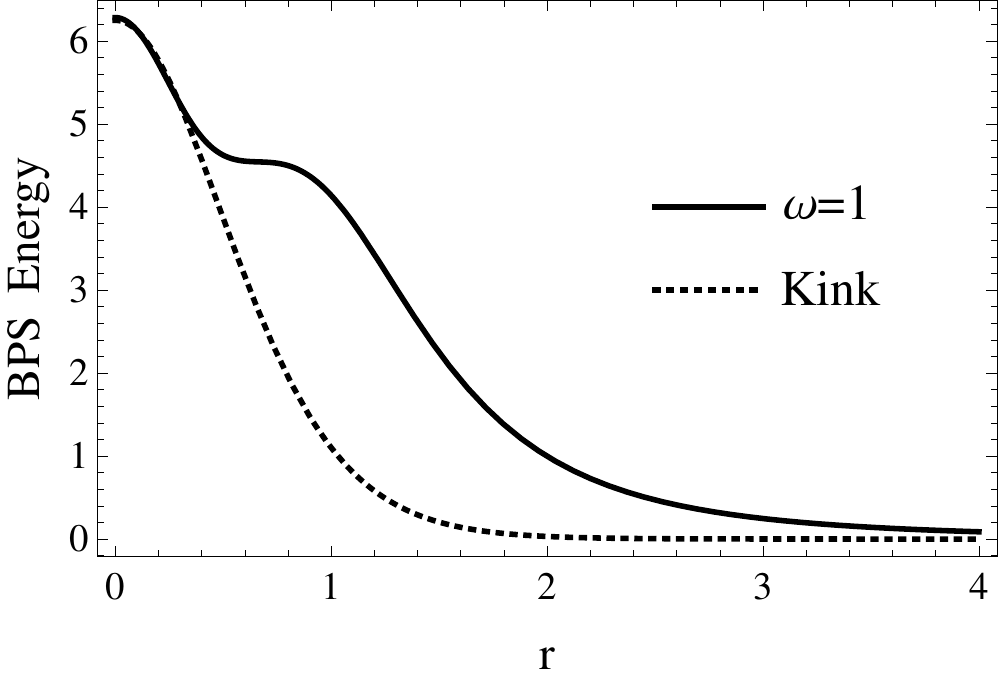}
\caption{BPS energy of the variable fields with $f(r=0)=0$ and $f(r\rightarrow\infty) =\pi $.}
\label{fig-0}
\end{figure}

Therefore, Fig. \ref{fig-1} represents kink-like solutions for the bound that Ref. \cite{12} argues has no finite energy solutions. Here it is worth mentioning that despite the existence of  kink-like solution for the variable field, it is not possible to use topological arguments to discuss the stability of the solutions of the model in this contour. As proposed in Ref. \cite{17}, kink solutions present finite bell-shaped energy solutions (proportional to $\text{sech}^{4}(r)$). Hence, a graphical solution with finite energy is expected for this new solution. Further, using the equation (\ref{energy-bps}) to analyse the BPS energy of the model, we obtain the plot of the energy density presented in Fig. \ref{fig-0}.

\subsection{From Kink to Compacton-like solutions}

\begin{figure}[!htb]
\centering
\includegraphics[scale=0.85]{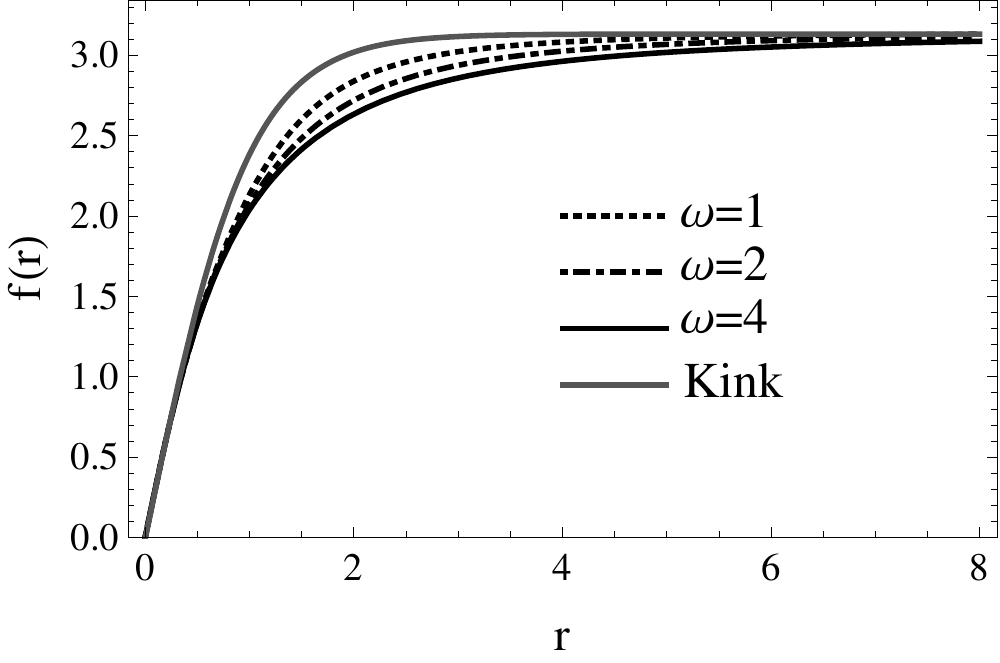}
\caption{Kink-like solutions for $f(r=0)=0$ and $f(r\rightarrow\infty)=\pi$.}
\label{fig-2}
\end{figure}

Since the kink-like solutions were found, let us now look for the compacton-like solutions in this model. For this, we need to analyse the parameter $\omega$. As a matter of fact, it appears in the expression (\ref{equationself}), and therefore it influences the dynamics of the variable field. With this in mind, we start our numerical investigation considering the variation of the parameter $\omega$ that is responsible for the contribution of the Chern-Simons field to the $O(3)$ sigma model. As a consequence,  the Fig. \ref{fig-2} was obtained.

\begin{figure}[!htb]
\centering
\includegraphics[scale=0.85]{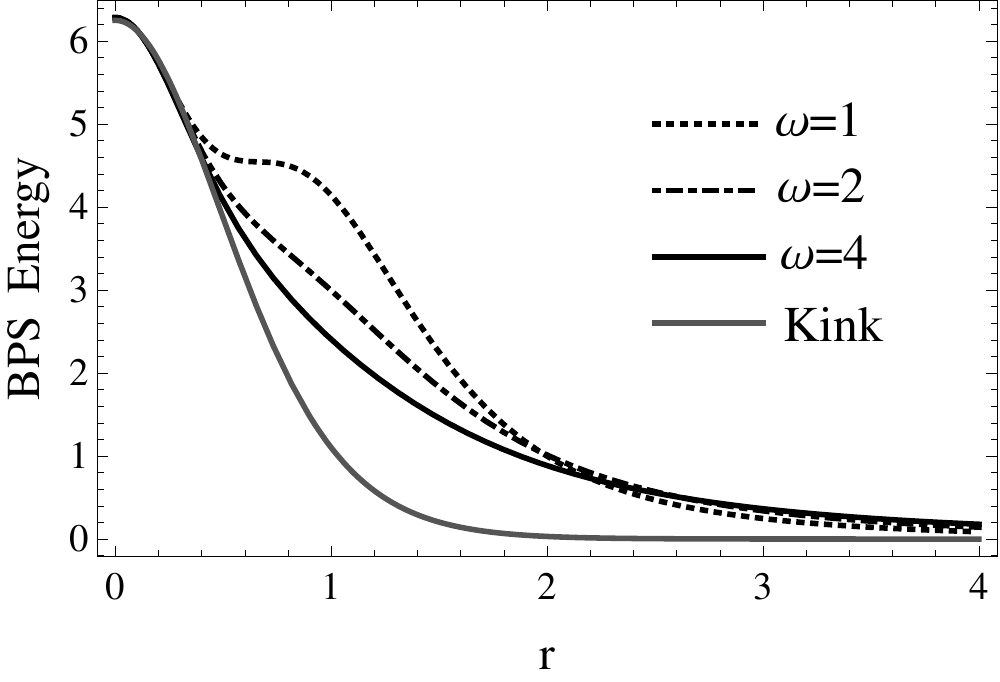}
\caption{BPS energy for $f(r=0)=0$ and $f(r\rightarrow\infty) =\pi $.}
\label{fig-3}
\end{figure}

Note from Fig. \ref{fig-2} that as the $\omega$ parameter decreases the solutions of the variable field tend to the solutions kink-like. In other words, as the contribution of the Chern-Simons term or the gauge field decreases the solutions of the variable field tend to a kink solutions. As consequence, the energy of the model assume a bell-shaped profile, as we can see from Fig. \ref{fig-3}.

\begin{figure}[!htb]
\centering
\includegraphics[scale=0.85]{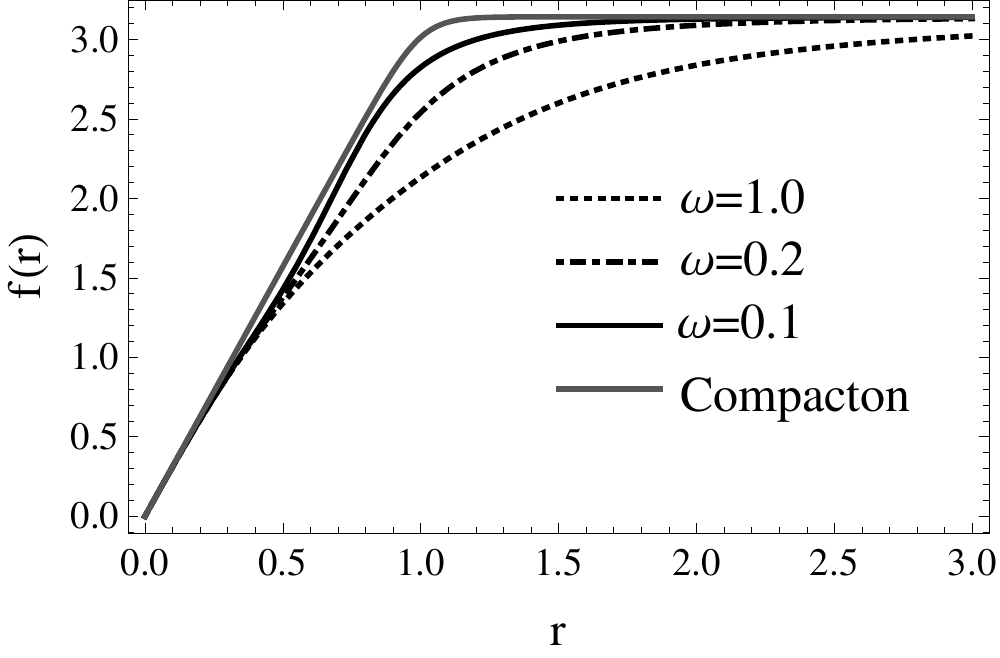}
\caption{From kink to compacton-like solutions of the model.}
\label{fig-5}
\end{figure}

According to the results in Fig. \ref{fig-3}, we note also that as the parameter $\omega$ decreases, the solution of the variable field takes the form of a kink solution, and the proper kink solution occurs when $\omega=0.35$. In order to transform from the kink-like solutions to compacton-like solutions, we continue to decrease the parameter $\omega$ until we achieved the compacton-like solutions as shown in Fig. \ref{fig-5}.

\begin{figure}[!htb]
\centering
\includegraphics[scale=0.85]{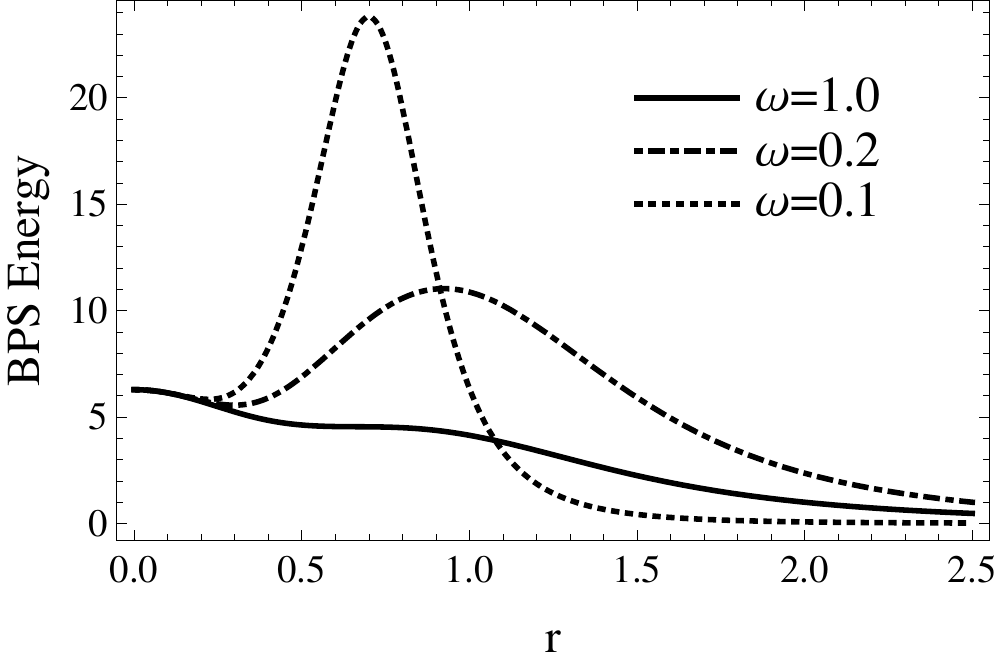}
\caption{BPS energy from the kink to compacton-like solutions of the model.}
\label{fig-6}
\end{figure}

Due to the characteristic of compacton-like solutions, we expect the BPS energy density to become more localized and sharp when kink solutions tend to compacton-like solutions, as presented in Fig. \ref{fig-6}.

\subsection{The existence of nontopological solitons}

A new numerical result appears when we assume that the $\omega$ parameter has a defined negative value. In this case, the kink-like and compact-like solutions will not exist. However, it is still possible to verify the solitary wave type solutions in the model, and in this case, we have nontopological solitons as solutions.

\begin{figure}[t]
\centering
\includegraphics[scale=0.85]{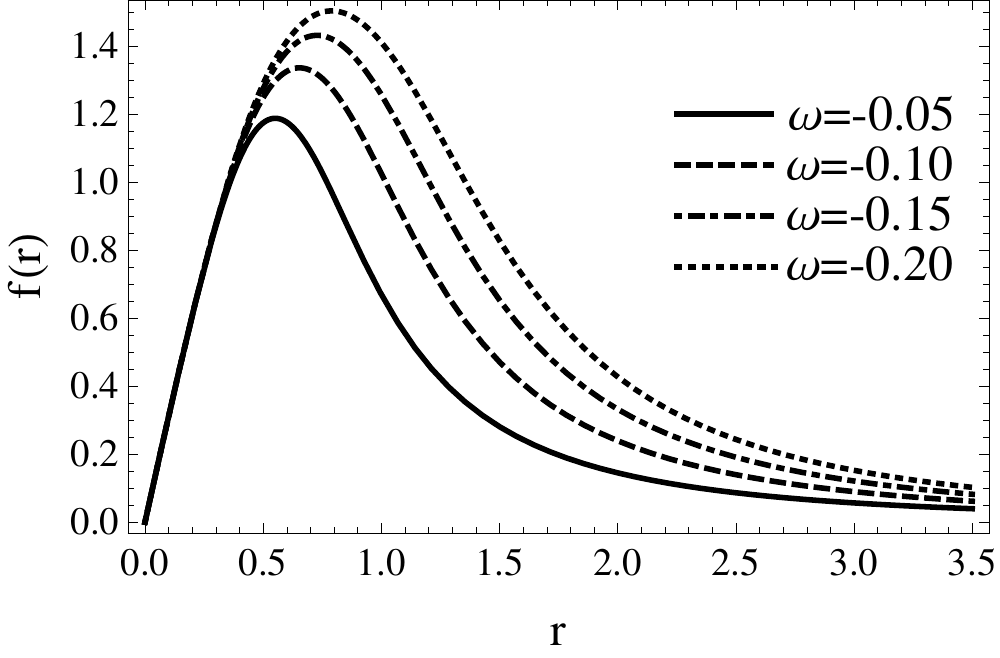}
\caption{Non-topological solitons for $\omega<0$.}
\label{fig-7}
\end{figure}

In this case, the variable field respects the following limit
\begin{align}
\lim_{r\rightarrow\infty}f(r)=0.
\end{align}

Due to this asymptotic behavior of the $f(r)$ variable field, we cannot use topological arguments to discuss the stability of solutions. Therefore, unlike topological solutions (kink-like and compacton-like solutions), to which the energy of topological solitons is given by $E=4\pi N$, the nontopological solutions do not have a quantized energy. 

\section{Concluding remarks}\label{section5}

In this work, we investigated the existence of solitonic solutions for the sigma-$O(3)$ model with generalized Chern-Simons term in the boundary $f(r=0)=0$. For this, we use a new parameter in the Chern-Simons term, which is also responsible for finitude of the energy density.  Hence, we found physically acceptable solutions limited by the boundary condition of the variable field $f(r=0)=0$ and $f(r\rightarrow\infty)=\pi$, in opposition to established literature, which claims the inexistence of solutions for this boundary condition \cite{12}. We note that when the parameter $\omega=1$, we fall into the case proposed by Ghosh \cite{12}. For $\omega>1$, we obtain kink-like solutions to the model with finite energy. Also, the parameter $\omega$ is responsible for transforming kink-type solutions into compacton-like solutions. Indeed, we observe that when the parameter $ 0 <\omega\leq 1$, we have kink-like solutions transforming into compacton-like solutions. As a consequence, the BPS energy density tends to become more localized and more sharper as the kink-like solutions tend to compacton-like. Physically, we can see this as a consequence of the magnetic induction being inversely proportional to this "compactification" parameter. 

We note that as we tend to compact solutions the contribution of the magnetic field increases significantly and in this way, the BPS energy density of the model will tend to a delta representation around of the singularity of the solution. Finally, we note that when $\omega$ is negative defined, there are no kink-like solutions, however, we can see the existence of non-topological solitons. As can be seen in Fig. (\ref{fig-7}), the wavelength of these structures becomes smaller as the parameter decreases. Similar results can be obtained for the boundary conditions studied in ref. \cite{12}.

\section*{Acknowledgements}
{The authors would like to thank the Funda\c{c}\~{a}o Cearense de Apoio ao Desenvolvimento Cient\'{i}fico  e  Tecnol\'{o}gico  (FUNCAP),  the  Coordena\c{c}\~{a}o  de  Aperfei\c{c}oamento  de  Pessoal  de N\'{i}vel Superior (CAPES) and the Conselho Nacional de Desenvolvimento Cient\'{i}fico e Tecnol\'{o}gico (CNPq) for financial support.}

\end{document}